# Laser Phase Noise Tolerance of Uniform and Probabilistically-shaped QAM Signals for High Spectral Efficiency Systems

Takeo Sasai, Asuka Matsushita, Masanori Nakamura, Seiji Okamoto, Fukutaro Hamaoka, and Yoshiaki Kisaka

(*Top-Scored Paper*)

*Abstract*—We numerically and experimentally investigate the laser phase noise tolerance of probabilistically shaped (PS) and uniformly shaped (US) quadrature amplitude modulation (QAM) signals. In the simulations, we compare PS-64QAM to US-16QAM, PS-256QAM to US-64QAM, and PS-1024QAM to US-256QAM under the same information rate (IR). We confirm that a sufficient shaping gain is observed with narrow linewidth lasers, whereas degradation of the shaping gain is clearly observed when large phase noise and high order modulation formats are assumed. In our experiments, we compare polarization-division-multiplexed (PDM) 16-GBd PS-1024QAM and US-256QAM under the same IR using lasers with 0.1-kHz and 40-kHz linewidths. For carrier phase recovery (CPR), we employ a pilot-assisted digital phase locked loop. Results reveal that PS-1024QAM achieves high performance with the 0.1 kHz-laser or > 5% pilot ratio, whereas US-256QAM outperforms PS-1024QAM when lasers with 40-kHz linewidth and < 5% pilot ratio are used. We also evaluate the pilot ratio dependency of the required optical signal-to-noise ratio at the forward error correction limit and the achievable information rate. Additionally, we compare the performance of two types of CPR updating schemes: updating phase estimation at only the pilot symbol or at all symbols.

## I. INTRODUCTION

To handle the ever-increasing data traffic in optical fiber communication, the continuous expansion of the fiber capacity is in demand. Both extending utilized bandwidth and improving spectral efficiency (SE) are necessary to expand the capacity. The latter is an effective approach to best utilize limited bandwidth resources and save costs per bit [1]. However, employing high order quadrature amplitude modulation (QAM) formats imposes strict requirements for maintaining a high optical signal-to-noise ratio (OSNR). Tremendous efforts have been put into improving *(i)* OSNR in transmission lines, including Raman amplification [2–4] and compensation of nonlinear impairments caused in fibers [5] and *(ii)* OSNR tolerance of transceivers (e.g., compensating the frequency responses of transceivers [6] and device nonlinearity [7]).

To further improve OSNR tolerance, a probabilistic shaping (PS) technique has attracted a lot of interest recently [8–11]. Fig. 1 illustrates constellation diagrams of uniformly-shaped (US) 16QAM and PS-64QAM under the same power and same information rate (IR). As shown in Fig. 1 (a), changing a constellation's symbol probability extends the symbol distance under a fixed signal power (PS-64QAM in this case), satisfying $d_U < d_P$, where $d_U$ denotes the symbol distance of US signals and $d_P$ denotes that of PS. Thus, PS signals have a higher tolerance of Gaussian noise than US lower order signals (US-16QAM in this case) under the same power and IR.

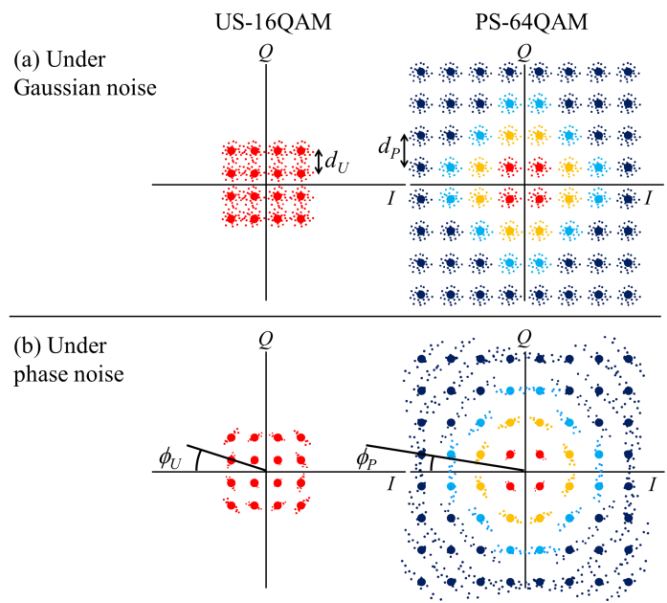

Fig. 1. US-16QAM and PS-64QAM constellation diagrams with (a) Gaussian noise and (b) phase noise under the same IR and the fixed signal power.

TABLE I
INFORMATION RATE AND ENTROPY OF PS AND US-QAM SIGNALS ASSUMING FEC CODE RATE OF 0.826

| | IR (bit/QAM symbol) | H (bit/QAM symbol) |
|---|---|---|
| US-16QAM | 3.305 | 4.000 |
| PS-64QAM | | 4.347 |
| US-64QAM | 4.958 | 6.000 |
| PS-256QAM | | 6.347 |
| US-256QAM | 6.611 | 8.000 |
| PS-1024QAM | | 8.347 |

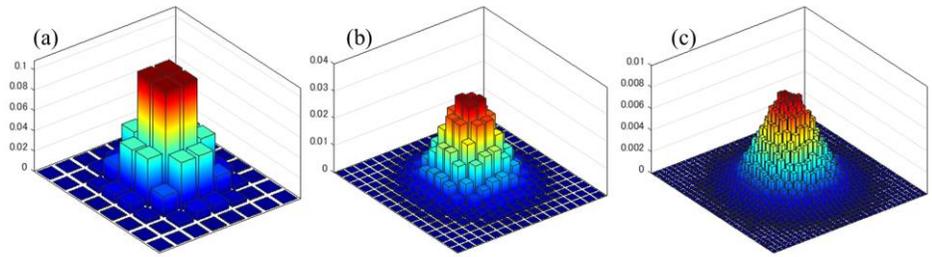

Fig. 2 Symbol probability distributions of PS- (a) 64QAM, (b) 256QAM, and (c) 1024QAM and the entropy $H$ are set to 4.347, 6.347, and 8.347, respectively, as listed in Table I.

However, PS signals inherently require higher order QAM than US signals under the same IR and the fixed code rate. That is, PS signals can be vulnerable to laser phase noise because their phase margin $\phi_P$ is smaller than the US signals' $\phi_U$ due to high multiplicity, as illustrated in Fig. 1 (b). Thus, it is unclear whether shaping gain can be sufficiently achieved in the presence of phase noise and the requirement for laser phase noise (i.e., laser linewidth) should be clarified for the design of future high SE systems. Several studies have been done on phase noise tolerance of US-QAM (< 64QAM) [12, 13]. Previously, we reported on experiments on the phase noise robustness of US-256QAM signals and revealed that laser linewidth (0.1~550 kHz) only slightly affected the signal quality [14]. A 48-GBd US-256QAM transmission was also demonstrated using a ~100-kHz laser [15]. For PS signals, the impact of PS on carrier phase recovery (CPR) performances is numerically analyzed in [16], and detuning the shaping factor of PS signals has been shown to yield stable CPR performance [17]. Moreover, nonlinear phase noise effects on PS signals due to long-haul transmission have been investigated [18], and a mitigation technique has also been proposed [19]. However, studies on the impact *laser phase noise* has on PS signals have been limited to numerical investigations [20].

In this paper, we numerically and experimentally investigate the laser phase noise tolerance of PS and US signals and compare their performance under the same IR [21]. We employ a pilot-assisted digital phase locked loop (PLL) for CPR [22]. In the simulation, we compare PS-64QAM to US-16QAM, PS-256QAM to US-64QAM, and PS-1024QAM to US-256QAM at 16 GBd. As a performance metric, we obtain the required SNR at the forward error correction (FEC) limit, sweeping the linewidth from 0 to 50 kHz. Results show that the shaping gain of PS signals significantly degrades under large phase noise when high order QAM such as PS-1024QAM is employed. Also, we experimentally investigate polarization-division-multiplexed (PDM) 16-GBd PS-1024QAM and US-256QAM under the same IR. As these results are inherently dependent on the CPR algorithm, we also show the pilot ratio dependency of the required OSNR at the FEC limit and achievable information rate (AIR). Results reveal that US-256QAM outperforms PS-1024QAM when widely used lasers (40-kHz linewidth) and a lower pilot ratio (<

5%) are employed. The phase noise is estimated at all symbols, including non-pilot symbols in CPR. For non-pilot symbols, we used a decision-directed estimation algorithm. One may argue that wrong decisions can lead to additional noise and CPR performance deterioration in such an algorithm, and wrong decisions are what cause PS performance degradation. To clarify this point, we also compare the performances of two types of updating schemes: updating phase noise estimation at all symbols and only at pilot symbols. We find that estimating phase noise at all symbols with a decision-directed PLL achieves a higher performance.

The remainder of this paper is organized as follows. Section II discusses the simulation of the phase noise tolerance of PS- and US-QAM signals. Section III experimentally investigates the phase noise tolerance of PDM 16-GBd PS-1024QAM and US-256QAM. Section IV concludes the paper.

## II. SIMULATION EVALUATION OF PHASE NOISE TOLERANCE OF PS- AND US-QAM SIGNALS

In this section, we compare the phase noise tolerance of PS- and US- signals by simulation. Section II-A describes the system model we used in the simulation, including the algorithm for phase noise compensation. We discuss the results in Section II-B.

### A. System model

To evaluate the phase noise tolerance, we employ three pairs of PS and US signals that have the same IR between PS and US signals. The investigated signals are listed in Table I with the IRs and the entropies, assuming the FEC code rate $R_c$ is 0.826 (21% overhead) [23]. The entropy in bit/QAM symbol is calculated as [11]

$$H = IR + m(1-R_c), \quad (1)$$

where *IR* is the information rate and set the same as US signals (e.g., for PS-64QAM, $IR = m_{16QAM} \times R_c = 4 \times 0.826 = 3.305$ bit/QAM symbol) and $m$ denotes bits of base constellation. Fig. 2 (a), (b), and (c) show the symbol probability distribution of PS-64, 256, and 1024QAM, respectively. Probabilistic shaping of the symbols was realized by independently shaping amplitudes of I and Q components of the QAM signal using Maxwell-Boltzmann distributions, and tuning the shaping factor to realize the target entropy shown in Table I.

In the simulation, laser phase noise is modeled as a Wiener process as [24]

$$\phi(n) = \phi(n-1) + w(n), \quad (2)$$

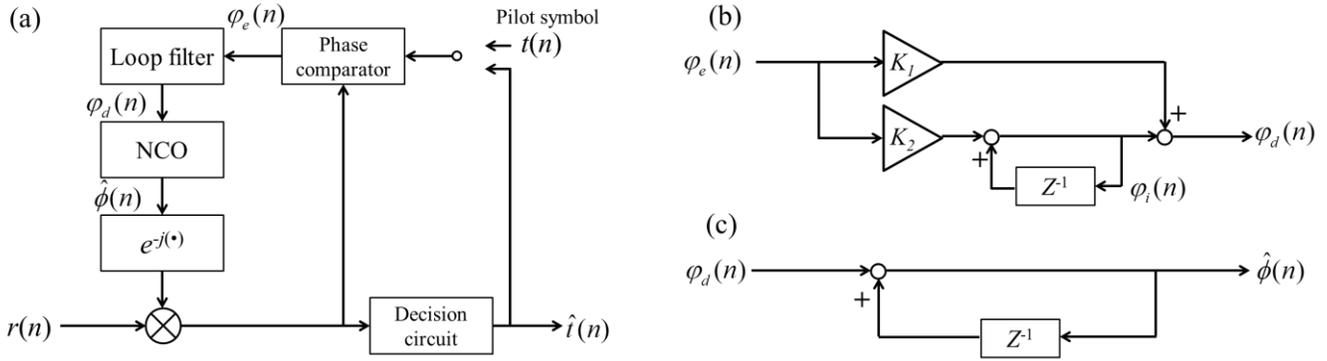

Fig. 3  Block diagrams of (a) the pilot-assisted digital PLL, (b) the loop filter in the PLL, and (c) the NCO in the PLL.

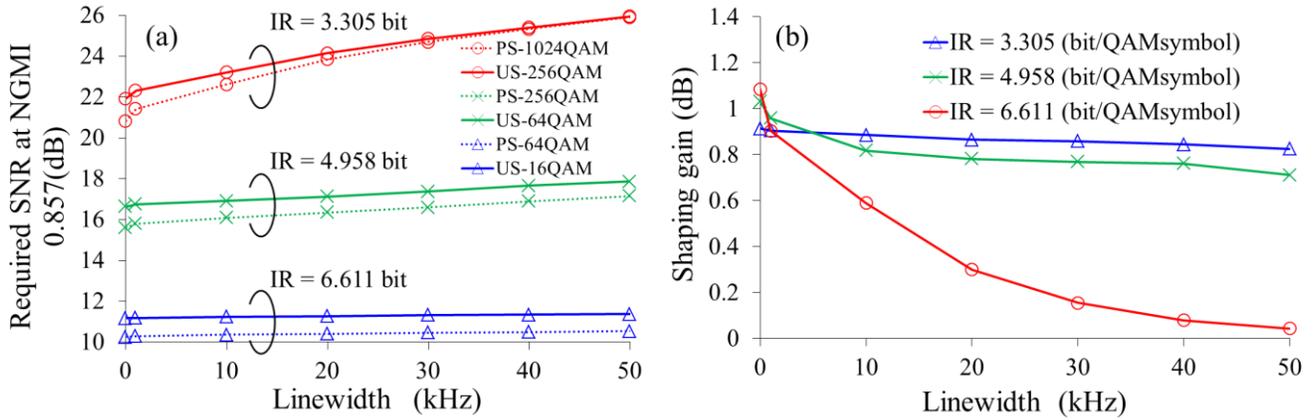

Fig. 4  Laser linewidth dependency of (a) required SNR at NGMI FEC limit 0.857 (21% overhead); (b) SNR gain due to PS.

where $\phi(n)$ is the phase noise of the $n$-th symbol and $w(n)$ is a Gaussian-distributed random variables with zero mean and variance

$$\sigma^2 = 2\pi \Delta f \tau , \quad (3)$$

where $\Delta f$ is the combined linewidth of transmitter and receiver lasers and $\tau$ is symbol duration. The incoming signals $r(n)$ is typically modeled as [25]

$$r(n) = t(n)e^{j\phi(n)} + N(n) , \quad (4)$$

where $t(n)$ is the $n$-th transmitted symbols and $N(n)$ is an additive white Gaussian noise.

In this simulation, a pilot-assisted digital PLL is employed for CPR [22]. Fig. 3 (a) shows the PLL algorithm we used. Note that the same algorithm is used in the experiments in Section III. $n$-th received signal $r(n)$ is compensated by estimated phase $\hat{\phi}(n)$ obtained by past symbols and is fed into a phase comparator (PC). When the symbol is a non-pilot symbol, the decision of the symbol is performed and the phase difference of the signal and the decision symbol $\hat{t}(n)$ is calculated in the PC as

$$\varphi_e(n) = \arg\left[r(n-1)e^{-j\hat{\phi}(n-1)}\right] - \arg\left[\hat{t}(n-1)\right], \quad (5)$$

where $\arg(\cdot)$ is the argument of a complex number. Note that the current symbol is used to estimate the phase noise of the next symbol. When the symbol is a pilot, $\hat{t}(n)$ is replaced by the pilot $t(n)$. Then, $\varphi_e(n)$ is filtered with a loop filter (see Fig. 3 (b)) and we obtain intermediate variables

$$\varphi_i(n) = \varphi_i(n-1) + K_1\varphi_e(n), \quad (6)$$

$$\varphi_d(n) = \varphi_i(n) + K_2\varphi_e(n), \quad (7)$$

where $K_1$ and $K_2$ are loop gains. Finally, the estimated phase noise is calculated via a numerically controlled oscillator (NCO) (see Fig. 3(c)) as

$$\hat{\phi}(n) = \hat{\phi}(n-1) + \varphi_d(n). \quad (8)$$

Before decoding, all the pilot symbols inserted into signals are eliminated. Then, the symbols are compared with the transmitted symbols and bit-wise log likelihood ratios (LLRs), generalized mutual information (GMI), and normalized GMI (NGMI) are calculated [27].

B.  Simulation results

The received data is generated according to Eq. (4) for different values of SNR and linewidth. The symbol rate is set to 16 GBd. In the simulation, PLL parameters $K_1$ and $K_2$ are optimized for each datum in the following results. The pilot ratio is set to 6.25%. For initial convergence of PLL, training symbols are inserted into first 100 symbols.

Fig. 4 (a) shows the required SNR of PS- and US-QAM signals at the NGMI limit 0.857 [23] as a function of linewidth $\Delta f$. Dashed lines represent PS signals, and solid lines represent US signals. Note that the noise component of the SNR in the vertical axis is calculated from additive white Gaussian noise $N(n)$ and thus phase noise impact can be observed as degradation of required SNR. The required SNR becomes higher as the IR (i.e., modulation order of QAM signals) becomes higher. Also, it seems that the shaping gain—defined as the difference between the required SNR of PS and US signals—remains constant for the IR of 3.305 and 4.958 bit (blue and green) regardless of linewidth, whereas the required SNR of PS-1024QAM (dashed red) gets closer to US-256QAM (solid red). This means PS-1024QAM is more vulnerable to the phase noise than US-256QAM is. Fig. 4 (b) shows the linewidth dependency of the shaping gain of each pair. When lower phase noise (< 1 kHz) is applied to the signals, expected shaping gain around 1 dB is clearly observed for all three pairs of signals. However, as the linewidth increases, the shaping gain significantly diminishes when PS-1024QAM is employed. The high order PS-QAM suffers from the phase noise impact and the benefit of PS diminishes.

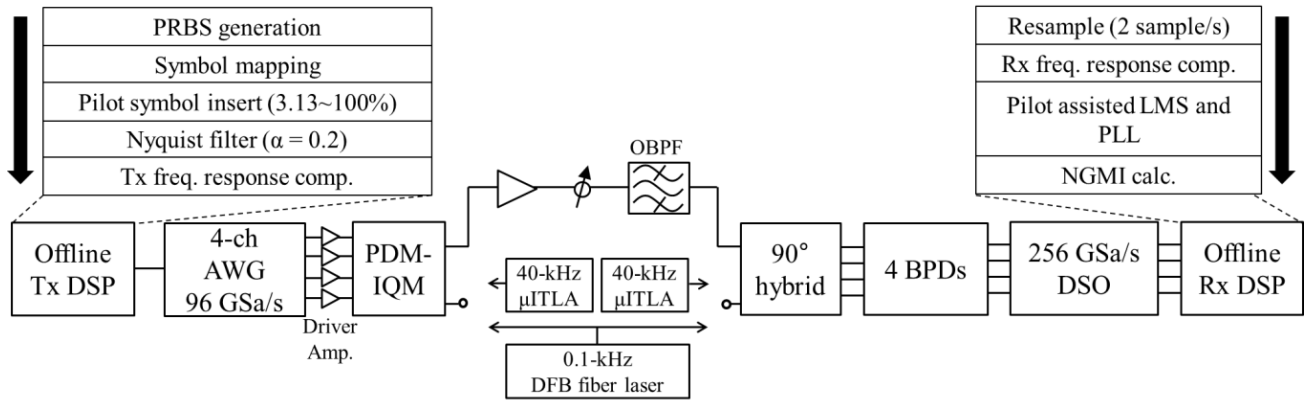

Fig. 5. Back-to-back experimental setup for evaluating PS-1024QAM and US-256QAM phase noise tolerance.

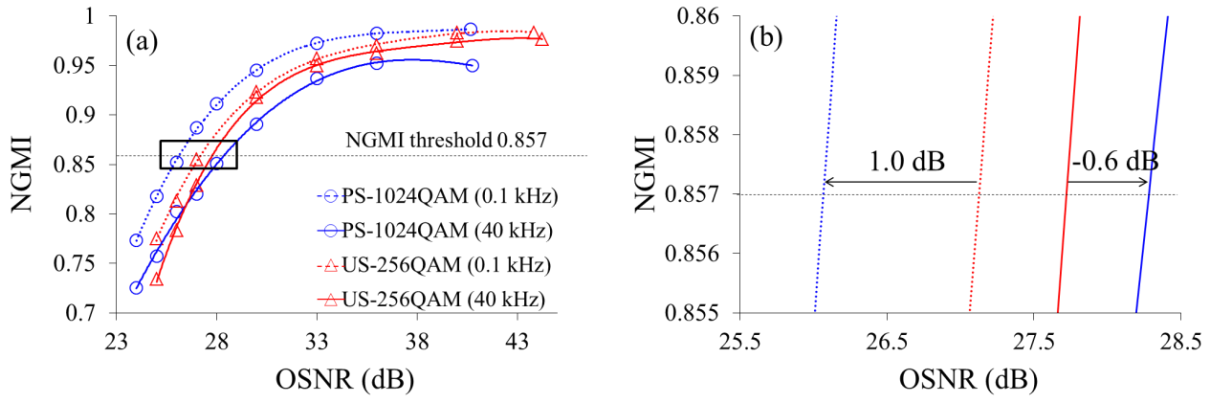

Fig. 6. (a) Back-to-back OSNR dependence of NGMI and (b) its enlarged version around 0.857 NGMI limit.

### III. Experimental Evaluation of Phase Noise Tolerance of PDM 16-GBd PS-1024QAM and US-256QAM

In this section, we experimentally investigate phase noise tolerance of PDM 16-GBd PS-1024QAM and US-256QAM. Section III-A describes the experimental setup and the OSNR characteristics of the signals. Section III-B shows the pilot ratio dependency of required OSNR and AIR. Section III-C discusses the impact of wrong decisions in a decision-directed PLL.

#### A. Experimental setup and OSNR characteristics

We prepared two types of lasers to evaluate the phase noise tolerance of signals. One was a micro-integrable tunable laser assemblies (µITLAs) laser with a 40-kHz linewidth. Beforehand, we experimentally examined their linewidth in the homodyne configuration with a delay line and a coherent receiver as Kikuchi proposed [26]. The other was a distributed feedback (DFB) fiber laser with a 0.1-kHz linewidth, which was used in a homodyne configuration as an ideal case reference with negligible phase noise and frequency offset (FO). To solely evaluate the impact of phase noise on signals, we performed precise FO compensation in receiver-side digital signal processing (DSP) when the 40-kHz lasers are used. This enabled us to fairly evaluate the impact of phase noise and compare the OSNR characteristics between the FO-compensated 40-kHz intradyne configuration and the 0.1-kHz homodyne configuration.

Using these lasers, we conducted back-to-back experiments for both PDM 16-GBd PS-1024QAM and PDM 16-GBd US-256QAM, as shown in Fig. 5. In the transmitter-side DSP, a random bit sequence was generated and mapped into PS-1024QAM or US-256QAM symbols as generated. For PS-1024QAM, a constant composition distribution matcher shaped the symbol probability distribution into a discrete Maxwell-Boltzmann distribution with 8.347 bit entropy assuming an FEC code rate of 0.826 (21% overhead) [23]. The IR was 6.611 bit/QAM symbols/pol for both signals. Then, pilot symbols were inserted for the adaptive equalizer and the CPR and their rates were arbitrarily changed (3.13 to 100%). Signals were Nyquist-pulse shaped by a root-raised-cosine filter with a 0.2 roll-off factor. The frequency response of the transmitter and the receiver was obtained in advance and compensated so that the desired signals were generated [6]. Electrical signals were emitted from an arbitrary waveform generator (AWG) with a 32-GHz analog bandwidth after being resampled to 96 Gsample/s. A PDM-IQ modulator (PDM-IQM) driven by 35-GHz driver amplifiers modulated the signals. The laser wavelength was set to 1555.752 nm.

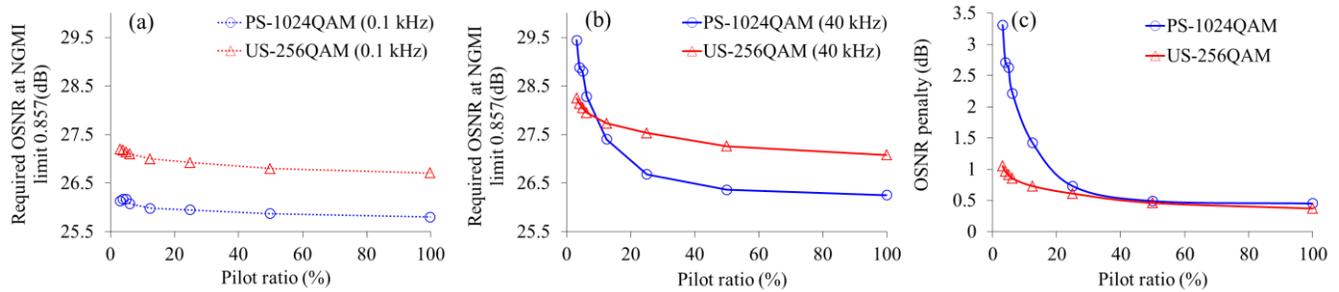

Fig. 7. Required OSNR at NGMI threshold of 0.857 versus pilot ratio for PS-1024QAM (blue) and US-256QAM (red) in the case of (a) 0.1 kHz and (b) 40 kHz laser. (c) Pilot ratio dependence of OSNR penalty due to laser phase noise.

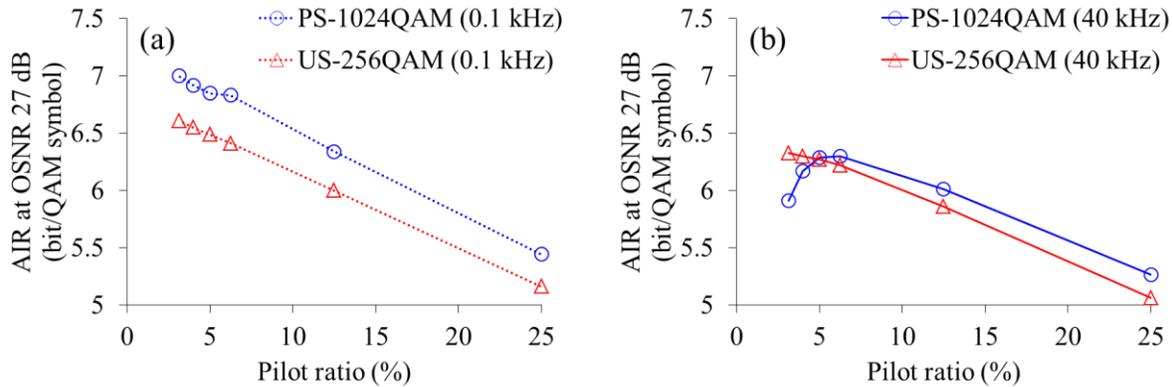

Fig. 8. AIR versus pilot ratio at OSNR 27dB for PS-1024QAM (blue) and US-256QAM (red) for (a) 0.1 kHz and (b) 40 kHz lasers.

In the optical domain, the signals were amplified with an erbium-doped fiber amplifier (EDFA) and filtered by an optical bandpass filter (OBPF) with a 1.6-nm bandwidth. A coherent receiver with a 90° hybrid and 70-GHz bandwidth balanced photo detectors (BPDs) converted the optical signals to electrical ones. The signals were digitized with a 256 Gsample/s digital sampling oscilloscope (DSO) with a 110-GHz bandwidth and fed into a receiver-side DSP. In the DSP, the signals were first resampled to 2 samples/symbol. The receiver-side frequency response was compensated using a finite impulse response filter, then polarization demultiplexing was performed with a 2 × 2 adaptive filter employing the least mean square algorithm (LMS). Phase noise was compensated, and the FO in a 40-kHz intradyne configuration was estimated by the pilot-assisted digital PLL introduced in Section II. In the 40-kHz intradyne configuration, we repeated the demodulation process and precisely compensated the FO before the adaptive filter, using the estimated phase noise in the PLL. The carrier phase was estimated at all symbols, and the PLL worked as a decision-directed type at non-pilot symbols. A decision-directed PLL could affect the compensation performance, and this point is discussed in detail in Section III-C. After demodulation, all the inserted pilot symbols were eliminated, and bit-wise LLRs, GMI, and NGMI were calculated [27].

We measured the back-to-back performance using both the 40-kHz lasers and the 0.1-kHz laser. For each laser, we obtained the OSNR characteristics of PS-1024QAM and US-256QAM, as shown in Fig. 6 (a). Fig. 6 (b) shows its enlarged version around the 0.857 pre-FEC NGMI limit [23]. Dashed lines represent 0.1-kHz homodyne and solid lines represent 40-kHz intradyne. In this experiment, the pilot ratio for the digital PLL was set to 6.25%. Regarding the 0.1-kHz laser, PS-1024QAM (blue) outperformed US-256QAM (red) at any OSNR, which means the shaping gain was acquired as expected because the phase noise impact was almost negligible. At the NGMI limit in particular, the shaping gain from US-256QAM was approximately 1.0 dB. However, in the 40-kHz case, no shaping gain was observed; even US-256QAM showed a higher OSNR tolerance than PS-1024QAM by 0.6 dB at the FEC limit. That is, the impact of phase noise was observed when practical 40-kHz lasers were used. In the lower OSNR region (< 26.5 dB), OSNR tolerance of PS-1024QAM was higher than US-256QAM in the 40-kHz case. This result is reasonable because lower OSNR means Gaussian noise (i.e., amplified spontaneous emission noise) is dominant and PS can present shaping gain in that region. In the high-OSNR region, phase noise is dominant over Gaussian noise, worsening PS performance.

### B. Pilot ratio dependency of required OSNR and AIR

The results shown in Section III-A depend on CPR performance because expected PS gain will be obtained when phase noise is sufficiently compensated. Thus, we evaluated the pilot ratio dependency of the OSNR required to achieve the FEC threshold.

Fig. 7 (a) and (b) show the required OSNR at NGMI limit 0.857 as a function of the pilot ratio for a 0.1-kHz laser and a 40-kHz laser, respectively. When a 0.1-kHz laser is employed, PS-1024QAM (dashed blue) requires a lower OSNR at any pilot ratio, and a sufficient shaping gain of around 1.0 dB exists. However, 40-kHz laser phase noise has a significant impact on PS-1024QAM at a low pilot ratio, while PS gain is clearly observed in the high pilot rate region. Fig. 7 (c) shows the pilot ratio dependency of the OSNR penalty at the NGMI limit. At a low pilot rate, US-256QAM is more tolerant of laser phase noise than PS-1024QAM. These results indicate that either narrow linewidth lasers or a CPR with a high pilot ratio have to be employed to acquire the PS gain in high SE modulation formats such as PS-1024QAM. However, the increase of pilot symbols in the data sequence is a trade-off with AIR. Therefore, we calculated AIR at the OSNR of 27 dB assuming ideal FEC, which is defined as

$$AIR = (1-r) \cdot GMI, \qquad (9)$$

where $r$ ($0 < r < 1$) is the pilot ratio and *GMI* is generalized mutual information [27]. Fig. 8 (a) and (b) correspond to the pilot ratio dependency of AIR for 0.1-kHz and 40-kHz linewidth, respectively. When the 0.1-kHz laser was used, AIR of PS-1024QAM outperformed that of US-256QAM at any pilot ratio. When the 40-kHz lasers were employed, the PS signals could achieve higher IRs at pilot ratio > 5%, while US signals showed higher AIR than PS signals in other regions. Maximum AIRs of both PS (at 7% pilot ratio) and US signals (at 3.13% pilot ratio) are almost the same, and determining which modulation format to choose is difficult.

These results indicate the choice of shaping form largely depends on the lasers and CPR pilot ratio employed in the system. If sufficiently narrow linewidth lasers can be made commercially available for use in future systems, the PS format will perform better due to its shaping gain. If the phase noise is large enough to deteriorate high-order QAM PS signals, CPR has to be improved or US signals utilized. Specifically, if the CPR is based on a pilot-assisted PLL, the pilot ratio should be carefully chosen so that the AIR becomes higher.

### C. CPR performances of estimating phase noise at all symbols and only at pilot symbols

The experiments we conducted were based on a pilot-assisted digital PLL, and it works as a decision-directed PLL at symbols other than pilot, as illustrated in Fig. 9 (a). One could argue that PS performance deterioration is simply due to additional noise arising from wrong decisions made at non-pilot symbols. To make this point clear, we compared CPR performance between updating phase estimation at all symbols and updating only at pilot symbols. Fig. 9 (b) corresponds to the latter. To evaluate the performance, we calculated the GMI difference between two schemes as

$$\Delta GMI = GMI_{(a)} - GMI_{(b)}, \qquad (10)$$

where $GMI_{(a)}$ denotes GMI when updating CPR at all symbols and $GMI_{(b)}$ is when updating CPR at only pilot symbols, corresponding to Fig. 9 (a) and (b), respectively. The pilot ratio dependency of *ΔGMI* at 27 dB OSNR is shown in Fig. 10. At any pilot ratio, updating phase estimation at all symbols yields a higher performance for both lasers. Furthermore, *ΔGMI* becomes larger as the pilot ratio becomes lower. This tendency is more remarkable when 40-kHz linewidth lasers are used than when 0.1-kHz lasers used. This can be explained by the fact that phase tracking phase noise of the 0.1-kHz linewidth is relatively easy for both updating schemes because its phase noise is almost negligible. The results indicate that phase noise tracking performance is insufficient in pilot-only updating. Rather than trying to reduce the decision-directed estimation, tracking phase noise at every symbol offers higher performances. One reason is that the symbol error rate and the wrong decision frequency are low enough to correctly estimate the carrier phase even when a decision-directed algorithm is employed. Secondly, updating at every symbol also provides higher tracking speed for phase noise. Consequently, all-symbol updating with the support of a decision-directed estimation is preferable even for high-order modulation formats such as PS-1024QAM.

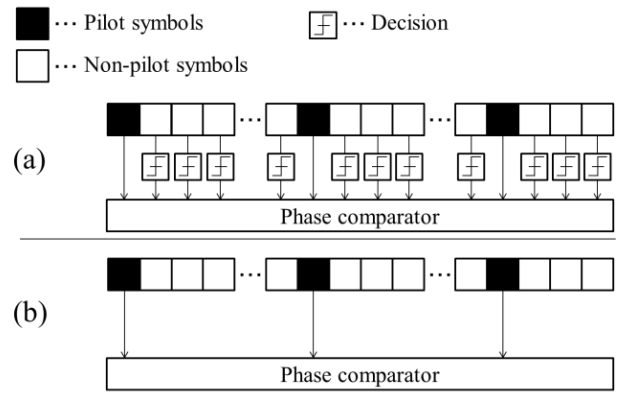

Fig. 9.  Phase noise estimation schemes at CPR. (a) Updating phase estimation at all symbols and (b) only at pilot symbols.

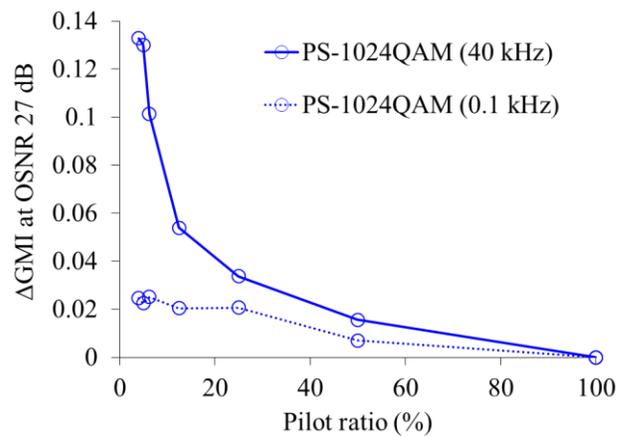

Fig. 10.  Pilot ratio dependency of ΔGMI (= GMI$_{(a)}$ − GMI$_{(b)}$). Updating at all symbols provides higher GMI than updating only at pilot symbols.

From these results, we conclude that the performance degradation of PS observed in Section III-A and III-B is due to laser phase noise itself rather than wrong decisions caused in the decision-directed PLL.

### IV. CONCLUSION

We numerically and experimentally investigated phase noise tolerances of high order PS- and US-QAM signals. In the simulations, we examined three pairs of PS and US signals with the same IR, PS-64QAM and US-16QAM, PS-256QAM and US-64QAM, and PS-1024QAM and US-256QAM. Simulation results show that the shaping gain diminishes when large phase

This is the author's version of an article that has been published in this journal. Changes were made to this version by the publisher prior to publication.
The final version of record is available at    http://dx.doi.org/10.1109/JLT.2019.2945470noise lasers and high-order QAM signal are employed. In the experiments, we evaluated PDM 16-GBd PS-1024QAM and US-256QAM using a 0.1-kHz self-homodyne laser and 40-kHz intradyne lasers. We obtained pilot ratio dependency of the required OSNR at a 21% FEC limit and AIR for each constellation. We also evaluated how the performance differed between updating phase estimation at all symbols and only at pilot symbols. The results show that PS performance can be worse than US signals under certain conditions in the experiments. For lasers with a sufficiently narrow linewidth, we achieved a high PS performance regardless of the pilot ratio. However, when we used practical lasers with a ~100-kHz linewidth, US-256QAM showed a higher OSNR tolerance and AIR than PS-1024QAM at a low pilot ratio (< 5%). Thus, we should take the laser linewidth and CPR algorithms into account when choosing a signal shaping form for designing a future system.

## REFERENCES

[1] S. L. Olsson *et al.*, "Record-high 17.3-bit/s/Hz spectral efficiency transmission over 50 km using probabilistically shaped PDM 4096-QAM," in *Proc. Opt. Fiber Commun. Conf. Expo.*, San Diego, CA, USA, Mar. 2018, Paper Th4C.5.

[2] X. Zhou *et al.*, "4000km transmission of 50GHz spaced, 10×494.85-Gb/s hybrid 32–64QAM using cascaded equalization and training-assisted phase recovery," in *Proc. Opt. Fiber Commun. Conf. Expo.*, Los Angeles, CA, USA, Mar. 2012, Paper PDP5C.6.

[3] M. Ionescu *et al.*, "91 nm C+L hybrid distributed Raman–erbium-doped fibre amplifier for high capacity subsea transmission," in *Proc. Eur. Conf. on Opt. Commun.*, Rome, Italy, Sep. 2018, Paper Mo4G.2.

[4] T. Kobayashi *et al.*, "PDM-16QAM WDM transmission with 2nd-order forward-pumped distributed Raman amplification utilizing incoherent pumping," in *Proc. Opt. Fiber Commun. Conf. Expo.*, San Diego, CA, USA, Mar. 2019, Paper Tu3F.6.

[5] D. Rafique *et al.*, "Compensation of intra-channel nonlinear fibre impairments using simplified digital back-propagation algorithm," *Opt. Express*, vol. 19, no. 10, pp. 9453–9460, May 2011.

[6] A. Matsushita *et al.*, "High-spectral-efficiency 600-Gbps/carrier transmission using PDM-256QAM format," *J. Lightw. Technol.*, vol. 37, no. 2, pp. 470–476, Jan. 2019.

[7] P.W. Berenguer *et al.*, "Nonlinear digital pre-distortion of transmitter components," in *Proc. Eur. Conf. Opt. Commun.*, Valencia, Spain, Sep. 2015, Paper Th.2.6.3.

[8] G. Bocherer *et al.*, "Bandwidth efficient and rate-matched low-density parity-check coded modulation," *IEEE Trans. Commun.*, vol. 63, no.12, pp. 4651–4665, Dec. 2015.

[9] F. Buchali *et al.*, "Experimental demonstration of capacity increase and rate-adaptation by probabilistically shaped 64-QAM," in *Proc. Eur. Conf. Opt. Commun.*, Valencia, Spain, Sep. 2015, Paper PDP.3.4.

[10] S. Chandrasekhar *et al.*, "High-spectral-efficiency transmission of PDM 256-QAM with parallel probabilistic shaping at record rate-reach trade-offs," in *Proc. Eur. Conf. Opt. Commun.*, Dusseldorf, Germany, Sep. 2016, Paper Th.3.C.

[11] J. Cho *et al.*, "Probabilistic constellation shaping for optical fiber communication," *J. Lightw. Technol.*, vol. 37, no. 6, pp. 1590–1607, March, 2019.

[12] M. Seimetz, "Laser linewidth limitations for optical systems with high-order modulation employing feed forward digital carrier phase estimation," in *Proc. OFC/NFOEC*, San Diego, CA, USA, Feb. 2008, Paper OTuM2.

[13] A. M. Ragheb *et al.*, "Laser phase noise impact on optical DP-MQAM: experimental investigation," *Photon. Netw. Commun.*, vol. 35, no. 2, pp. 237–244, Apr. 2018

[14] T. Sasai *et al.*, "Experimental investigation of laser linewidth tolerance of 32-GBaud DP-256QAM optical coherent system," in *Proc. OECC*, Jeju, South Korea, Jul. 2018, Paper 6D2-1.

[15] A. Matsushita *et al.*, "Single-carrier 48-GBaud PDM-256QAM transmission over unrepeated 100 km pure-silica-core fiber using commercially available μITLA and LN-IQ-modulator," in *Proc. Opt. Fiber Commun. Conf. Expo.*, San Diego, CA, USA, Mar. 2018, Paper M1C1.

[16] D. A. A. Mello *et al.*, "Interplay of probabilistic shaping and the blind phase search algorithm," *J. Lightw. Technol*, vol. 36, no. 22, pp. 5096–5105, Nov. 2018.

[17] F. A. Barbosa *et al.*, "Shaping factor detuning for optimized phase recovery in probabilistically-shaped systems," in *Proc. Opt. Fiber Commun. Conf. Expo.*, San Diego, CA, USA, Mar. 2019, Paper W1D.4.

[18] D. Pilori *et al.*, "Comparison of probabilistically shaped 64QAM with lower cardinality uniform constellations in long-haul optical systems," *J. Lightw. Technol*, vol. 36, no. 2, pp. 501–509, Jan. 2018.

[19] D. Pilori *et al.*, "Non-linear phase noise mitigation over systems using constellation shaping," *J. Lightw. Technol*, vol. 37, no. 14, pp. 3475–3482, Mar. 2019.

[20] S. Okamoto *et al.*, "Laser phase noise tolerance of probabilistically-shaped constellations" in *Proc. Opt. Fiber Commun. Conf. Expo.*, San Diego, CA, USA, Mar. 2018, Paper W2A.51.

[21] T. Sasai *et al.*, "Experimental analysis of laser phase noise tolerance of uniform 256QAM and probabilistically shaped 1024QAM," in *Proc. Opt. Fiber Commun. Conf. Expo.*, San Diego, CA, USA, Mar. 2019, Paper W1D.5.

[22] T. Kobayashi *et al.*, "160-Gb/s polarization-multiplexed 16-QAM long-haul transmission over 3,123 km using digital coherent receiver with digital PLL based frequency offset compensator," in *Proc. OFC/NFOEC*, San Diego, CA, USA, Mar. 2010, Paper OTuD1.

[23] M. Nakamura *et al.*, "Spectrally efficient 800 Gbps/carrier WDM transmission with 100-GHz spacing using probabilistically shaped PDM-256QAM," in *Proc. Eur. Conf. Opt. Commun.*, Rome, Italy, Sep. 2018, Paper We3G.5.

[24] T. Pfau *et al.*, "Hardware-efficient coherent digital receiver concept with feedforward carrier recovery for M-QAM constellations," *J. Lightw. Technol*, vol. 27, no. 8, pp. 989–999, Apr. 2009.

[25] I. Fatadin *et al.*, "Laser linewidth tolerance for 16-QAM coherent optical systems using QPSK partitioning," *IEEE Photon. Technol. Lett.*, vol. 22, no. 9, pp. 631–633, May. 1, 2010.

[26] K. Kikuchi, "Characterization of semiconductor-laser phase noise and estimation of bit-error rate performance with low-speed offline digital coherent receivers," *Opt. Express*, vol. 20, no. 5, pp. 5291–5302, Feb. 2012

[27] J. Cho *et al.*, "Normalized generalized mutual information as a forward error correction threshold for probabilistically shaped QAM," in *Proc. Eur. Conf. Opt. Commun.*, Gothenburg, Sweden, Sep. 2017, Paper M.2.D.2.**Takeo Sasai** received the B.E. and M.E. degrees in Electrical Engineering and Information Systems from the University of Tokyo, Tokyo, Japan, in 2014 and 2016, respectively. In 2016, he joined NTT Network Innovation Laboratories, Yokosuka, Japan. He has been engaged in R&D of digital coherent optical transmission systems. His research interests include high-capacity optical transport systems with digital signal processing and machine learning.

**Asuka Matsushita** received the B.E. and M.E. degrees in Electric and Photonic Systems from Waseda University, Tokyo, Japan, in 2012 and 2014, respectively. In 2014, she joined NTT Network Innovation Laboratories, Yokosuka, Japan. Her research interests include high-capacity optical transport systems with digital signal processing. She is a member of the Institute of Electronics, Information, and Communication Engineers, Tokyo, Japan.

**Masanori Nakamura** received the B.S. and M.S. degrees in Applied Physics from Waseda University, Tokyo, Japan, in 2011 and 2013, respectively. In 2013, he joined NTT Network Innovation Laboratories, Yokosuka, Japan, where he engaged in research on high-capacity optical transport networks. He is a member of the Institute of Electronics, Information and Communication Engineers, Tokyo, Japan. He received the 2016 IEICE Communications Society Optical Communication Systems Young Researchers Award.Copyright (c) 2019 IEEE. Personal use is permitted. For any other purposes, permission must be obtained from the IEEE by emailing pubs-permissions@ieee.org.

**Seiji Okamoto** received the B.S., M.S., and Ph.D degrees in Electrical Engineering from Tohoku University, Sendai, Japan, in 2009, 2011, and 2018, respectively. In 2011, he joined NTT Network Innovation Laboratories, Yokosuka, Japan. He has been engaged in R&D of digital coherent optical transmission systems. His research interests include digital signal processing for low-power coherent DSP and wideband transmission. He is a member of the Institute of Electronics, Information and Communication Engineers.

**Fukutaro Hamaoka** received the B.E., M.E., and Ph.D. degrees in Electrical Engineering from Keio University, Yokohama, Japan, in 2005, 2006, and 2009, respectively. From 2009 to 2014, he was with NTT Network Service Systems Laboratories, Musashino, Japan, where he was engaged in R&D of high-speed optical communications systems, including digital coherent optical transmission systems. He is currently with NTT Network Innovation Laboratories, Yokosuka, Japan. His research interests include high-capacity optical transport systems with digital signal processing. He is a member of the Institute of Electronics, Information and Communication Engineers, Tokyo, Japan. In 2007, he received the Japan Society of Applied Physics Young Scientist Presentation Award.

**Yoshiaki Kisaka** received the B.E. and M.E. degrees in Physics from Ritsumeikan University, Kyoto, Japan, in 1996 and 1998, respectively. In 1998, he joined NTT Optical Network Systems Laboratories, Yokosuka, Kanagawa, Japan, where he was engaged in R&D of high-speed optical communication systems including 40-Gbit/s/ch WDM transmission systems and mapping/multiplexing schemes in optical transport networks (OTNs). From 2007 to 2010, he was with the NTT Electronics Technology Corporation, where he was engaged in the development of 40/100-Gbit/s OTN framer LSIs. His current research interests are high-speed and high-capacity optical transport networks that use digital coherent technology. Since 2010, he has contributed to R&D of digital coherent signal processors at 100 Gbit/s and beyond. He is a member of the Institute of Electronics, Information and Communication Engineers. He received the Young Engineer's Award from IEICE in 2001.